\documentclass[runningheads]{rnl}

\usepackage[utf8]{inputenc}      
\usepackage[frenchb]{babel}
\usepackage{color}

\usepackage{graphicx, subfigure}  

\usepackage{amsmath,amsfonts,amssymb}   

\begin{document}
 
\title*{Dynamiques transitoires de sillage dans le \og pinball fluidique\fg}
 
 
\titrecourt{Dynamiques transitoires de sillage dans le \og pinball fluidique\fg }

\author{Deng Nan\inst{1,2}
\and Pastur Luc R.\inst{1}
\and Noack Bernd R.\inst{2,3,4}
\and Cornejo-Maceda Guy\inst{2}
\and Lusseyran Fran\c{c}ois\inst{2}
\and Loiseau Jean-Christophe\inst{5}
\and Morzy\'nski Marek\inst{6}
}

\index{Deng Nan}              
\index{Pastur Luc R.}
\index{Noack Bernd R.}
\index{Cornejo-Maceda Guy}
\index{Lusseyran Franc{c}ois}
\index{Loiseau Jean-Christophe}
\index{Morzy\'nski Marek}

\auteurcourt{Deng \textit{et al}}
 
\adresse{IMSIA -- UMR9219 , ENSTA ParisTech, Palaiseau, France
\and LIMSI -- CNRS, Université Paris Saclay, Orsay, France
\and Harbin Institute of Technology, China
\and Tecnische Universit\"at Berlin, Allemagne 
\and Laboratoire DynFluid, \'Ecole Nationale Sup\'erieure d'Arts et M\'etiers, Paris, France
\and Poz\'nan University of Technology, Pologne
}

\email{nan.deng@ensta-paristech.fr}

\maketitle              

\begin{resume}
Nous nous intéressons à la dynamique transitoire d'une configuration fluide formée de trois cylindres fixes distribués sur un triangle équilatéral en écoulement transverse (\og pinball fluidique\fg), pour différentes valeurs du nombre de Reynolds sur la route vers le chaos. Nous étudions plus particulièrement l'action des degrés de liberté élémentaires du système dynamique sous-jacent sur les coefficients de trainée et de portance de l'écoulement fluide. 
\end{resume}

\begin{resumanglais}
In this work, we are interested in the transient dynamics of a fluid configuration consisting of three fixed cylinders whose axes distribute over an equilateral triangle in transverse flow (\og fluidic pinball \fg). As the Reynolds number is increased on the route to chaos, its transient dynamics tell us about the contribution of the elementary degrees of freedom of the system to the lift and drag coefficients. 
\end{resumanglais}

\section{Introduction}

La configuration d'écoulement dite du \og pinball fluidique\fg \ a été récemment introduite avec l'objectif de proposer un système à la fois simple et rapide à simuler numériquement  pour expérimenter différentes techniques de contrôle en mécanique des fluides, et suffisamment riche pour tester les problèmes liés aux entrées et sorties multiples dans ces systèmes, dont la dynamique est intrinsèquement non-linéaire et la dimension de l'espace des états virtuellement infinie (équations de Navier-Stokes). Il s'agit de trois cylindres disposés sur les sommets d'un triangle équilatéral en écoulement transverse, dont les actionneurs sont les cylindres eux-mêmes, susceptibles de tourner sur leur axe propre, tandis que les capteurs sont des sondes de vitesse ou de pression placées dans le sillage ou à la surface des cylindres \cite{deng:noack2017}. La dynamique \textit{naturelle}, non forcée, de cette configuration d'écoulement s'est révélée étonnamment riche \cite{deng:jfm2019}. C'est ce que nous souhaitons mettre en évidence dans cette contribution, où les dynamiques transitoires du système dynamique sous-jacent, étudiées du point de vue des coefficients de portance et de trainée du système fluide, sont instructives quant aux mécanismes à l'\oe uvre dans l'écoulement. C'est en particulier le cas vis-à-vis des deux bifurcations, Hopf puis fourche supercritiques, subies par le système, pour des nombres de Reynolds croissants, sur sa route vers le chaos.
 
\section{Configuration d'écoulement}

\begin{figure}
 \centerline{
 \includegraphics[width=.55\linewidth]{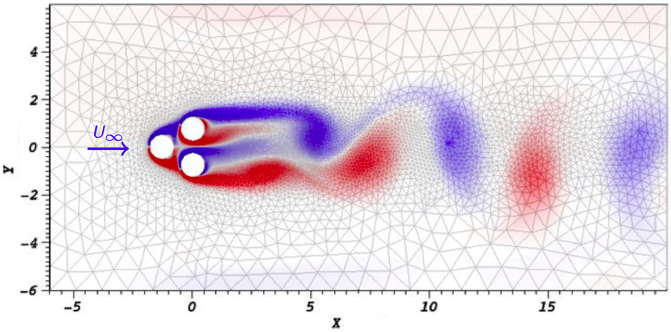}
 }
\caption{Configuration dite du \og pinball fluidique\fg \ et dimensions du domaine simulé. Un champ typique de vorticité est représenté en couleur. La vitesse amont est notée $U_\infty$.}
\label{dengfig:pinball}
\end{figure}

La configuration géométrique, bi-dimensionnelle dans le plan $(x,y)$, est constituée de trois cylindres fixes de diamètre $D$ montés sur les sommets d'un triangle équilatéral de côté $3D/2$, cf figure~\ref{dengfig:pinball}. L'écoulement amont, de vitesse uniforme $U_\infty$ en entrée du domaine, est transverse, en incidence nulle sur le cylindre amont. La résolution des équations de Navier-Stokes s'appuie sur une méthode de discrétisation en éléments finis du second ordre de type Taylor-Hood \cite{deng:taylor1973}, sur une grille non structurée de 4\,225 triangles et 8\,633 sommets, et une intégration implicite en temps du troisième ordre \cite{deng:noack2003,deng:noack2017}. Le nombre de Reynolds est défini par $Re=U_\infty D/\nu$, où $\nu $ est la viscosité cinématique du fluide. La condition de sortie est à contrainte nulle. Le coefficient de portance $C_L=2F_L/\rho U_\infty^2$ et le coefficient de trainée $C_D=2F_D/\rho U_\infty^2$  sont calculés à partir de la résultante des forces de pression et des forces visqueuses qui s'exercent sur les trois cylindres. 

\section{Route vers le chaos}

La route vers le chaos dans cette configuration fait apparaître une première bifurcation de Hopf supercritique pour $Re=Re_{HP}\approx 20$, associée au lâcher cyclique de tourbillons contra-rotatifs dans le sillage des trois cylindres à partir des couches cisaillées qui délimitent la configuration, formant une allée de tourbillons de von K\'arm\'an. Pour la valeur critique $Re=Re_{FC}\approx 70$, le système subit une bifurcation fourche supercritique associée à une brisure de symétrie du champ moyen, le jet entre les deux cylindres aval étant défléchi vers le haut ou le bas. A $Re=Re_{QP}\approx 100$, la dynamique devient quasi-périodique, avec des modulations basse fréquence de la position du jet. Enfin, pour $Re>Re_{CH} \approx 115$, la dynamique du système est pleinement chaotique et le champ moyen retrouve sa symétrie initiale \cite{deng:fedsm2018}. 

\begin{figure}
\centerline{
 \includegraphics[width=.32\linewidth]{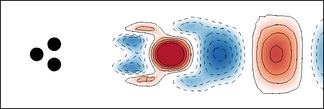} \hfill \includegraphics[width=.32\linewidth]{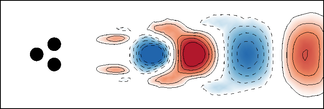} \hfill \includegraphics[width=.32\linewidth]{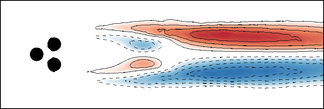} 
 }

\centerline{
 \includegraphics[width=.32\linewidth]{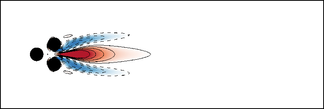} \quad \includegraphics[width=.32\linewidth]{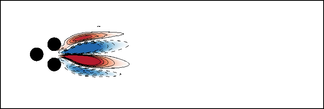} 
 }
\caption{Champs de vorticité des modes $\mathbf{u}_1(\mathbf{x})$, $\mathbf{u}_2(\mathbf{x})$, $\mathbf{u}_3(\mathbf{x})$ (en haut), $\mathbf{u}_4(\mathbf{x})$, $\mathbf{u}_5(\mathbf{x})$ (en bas), associés aux cinq degrés de liberté élémentaires $\left\{a_1(t) - a_5(t)\right\}$, à $Re=80$. }
\label{dengfig:ui}
\end{figure}

Deng \textit{et al} (2019) ont montré que la dynamique de l'écoulement fluide résultant de la bifurcation de Hopf supecritique 
%
%
pouvait être décrite par trois degrés de liberté $a_1$, $a_2$, $a_3$, associés aux modes $\mathbf{u}_1(\mathbf{x})$, $\mathbf{u}_2(\mathbf{x})$, $\mathbf{u}_3(\mathbf{x})$ de la figure~\ref{dengfig:ui}. Les deux premiers degrés de liberté $a_{1,2}(t)$ sont associés aux oscillations de l'écoulement et à l'allée de von K\'arm\'an. Le troisième degré de liberté $a_3(t)$ rend compte de la distorsion de l'état de base par les fluctuations $a_{1,2}$, via le tenseur de Reynolds, depuis l'état de base stationnaire symétrique instable jusqu'au champ moyen asymptotique \cite{deng:chaos2018,deng:loiseau2017}. La bifurcation fourche supercritique peut quant à elle être décrite par deux degrés de liberté supplémentaires $a_4$, $a_5$, associés aux modes $\mathbf{u}_4(\mathbf{x})$, $\mathbf{u}_5(\mathbf{x})$ de la figure~\ref{dengfig:ui}. Les degrés de liberté $a_4$, $a_5$, sont responsables de la brisure de symétrie induite par la bifurcation fourche. Les champs de vitesse associés aux degrés de liberté $a_3$ et $a_5$ sont symétriques, tandis que ceux associés à $a_1$, $a_2$ et $a_4$ sont anti-symétriques. Les modes $\left\{\mathbf{u}_i\right\}$ sont orthogonaux entre eux; tous les détails sur leur construction sont donnés dans \cite{deng:jfm2019}. Au seuil de la bifurcation fourche, la dynamique de l'écoulement peut être décrite par le modèle réduit minimal à cinq degrés de liberté \cite{deng:jfm2019}:
\begin{eqnarray}
 \dot{a}_1 & = & \sigma (a_3)a_1 + \omega (a_3)a_2 \label{dengeq:a1} \\
 \dot{a}_2 & = & \sigma (a_3)a_2 - \omega (a_3)a_1 \label{dengeq:a2} \\
 {a}_3 & = & \kappa (a_1^2+a_2^2) \label{dengeq:a3} \\
 \dot{a}_4 & = & \sigma _4a_4 - \beta _5a_5a_4 \\
 a_5 & = & \kappa_4a_4^2 \label{dengeq:a5} 
\end{eqnarray}
où $\sigma(a_3) = \sigma _1 - \beta a_3$, $\omega(a_3) = \omega _1 - \delta a_3$, $\omega _1$ étant la pulsation de l'allée de von K\'arm\'an au voisinage de l'état stationnaire instable. Tous les coefficients du système d'équations sont positifs. Les équations~\eqref{dengeq:a3},\eqref{dengeq:a5} traduisent la servitude des degrés de liberté $a_3$ et $a_5$ vis-à-vis des degrés de liberté dominants $a_{1,2}$ et $a_4$, respectivement. 
A plus haut nombre de Reynolds, les degrés de liberté se couplent et des termes croisés, linéaires et quadratiques du vecteur d'état $\mathbf{a}=(a_1\ a_2\ a_3\ a_4\ a_5)^t$, apparaissent à chacune des lignes du système \cite{deng:jfm2019}:
\begin{equation}
 \dot{a}_i = \sum_{j=1}^5 \ell_{ij} a_j + \sum_{j=1}^5 \sum_{k=1}^5 q_{ijk}a_ja_k.
\end{equation}

\section{Dynamiques transitoires et états asymptotiques}
\label{dengsec:4}

\begin{figure}
\centerline{
\begin{tabular}{c c c}
$Re$ & $C_D$ & $C_L$ \\
\hline \hline \vspace*{-3mm}
 80  & \raisebox{-0.92\height}{\includegraphics[width=.45\linewidth]{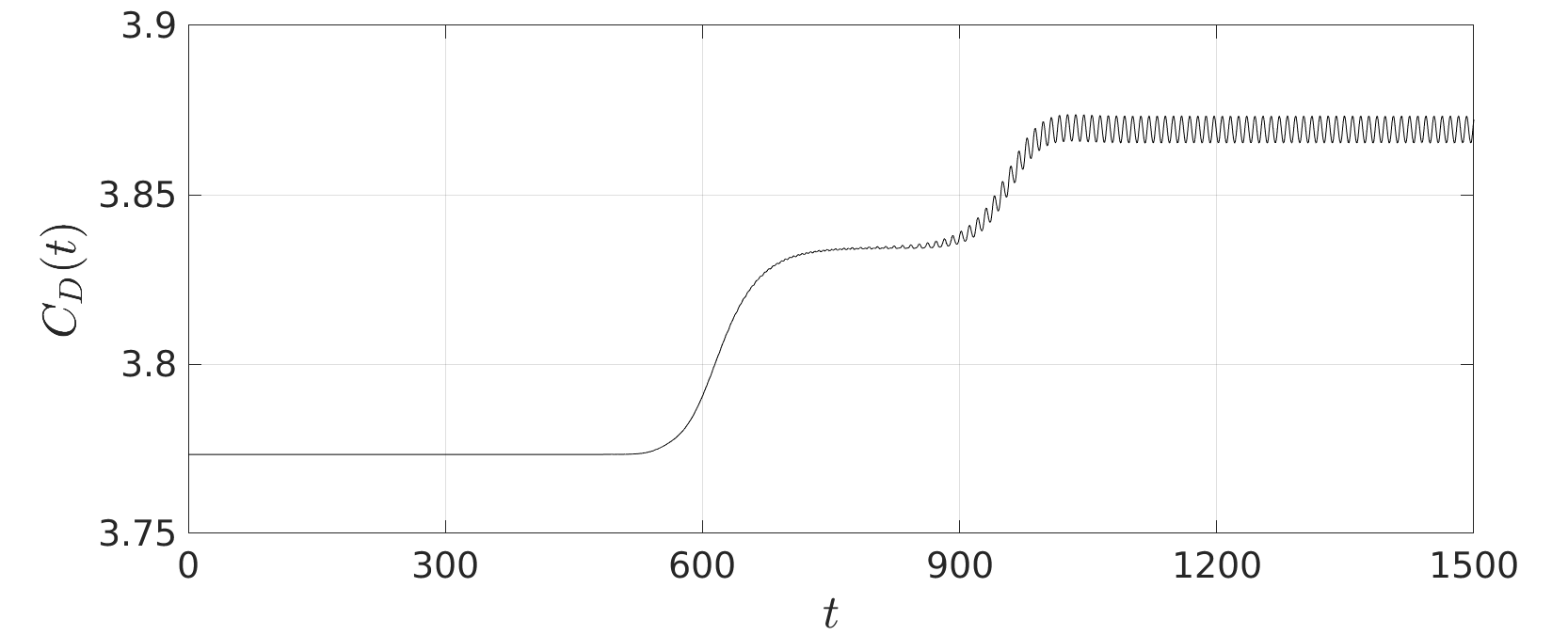}} & \raisebox{-0.92\height}{\includegraphics[width=.45\linewidth]{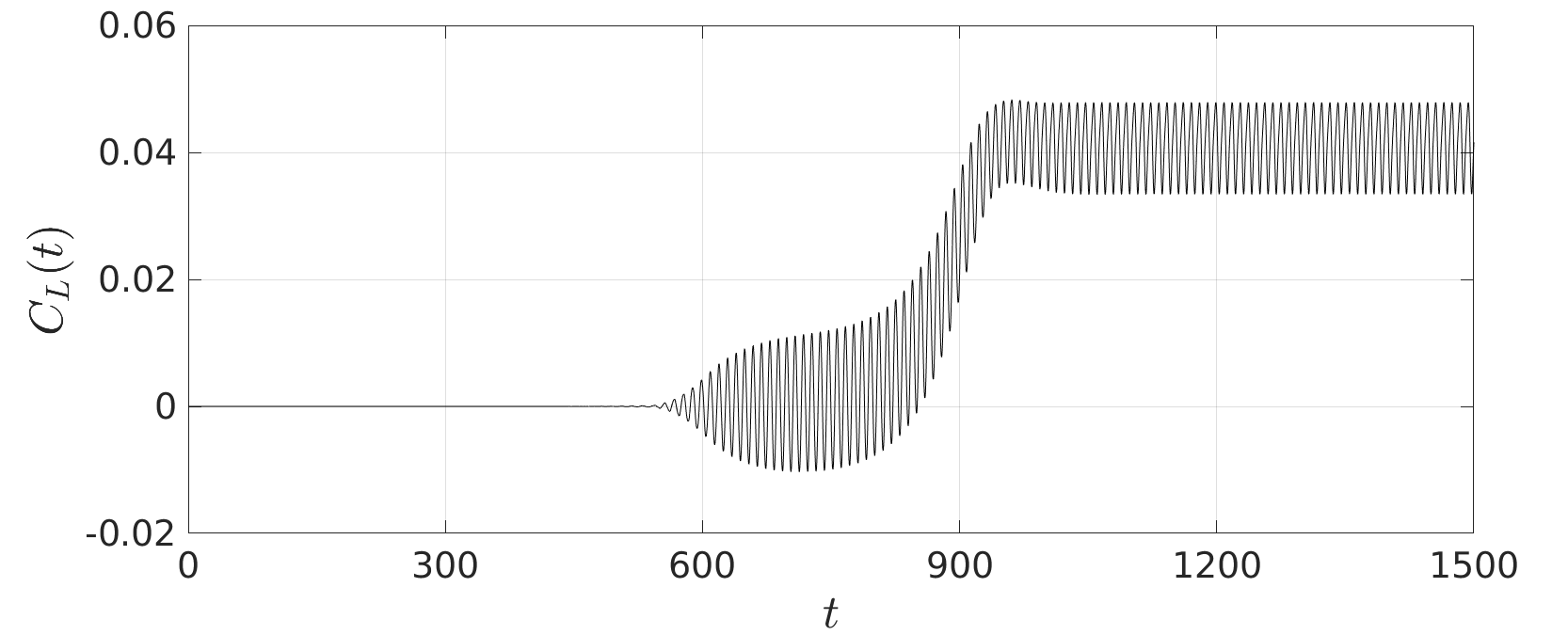}} \\ \vspace*{-3mm}
 100 & \raisebox{-0.92\height}{\includegraphics[width=.45\linewidth]{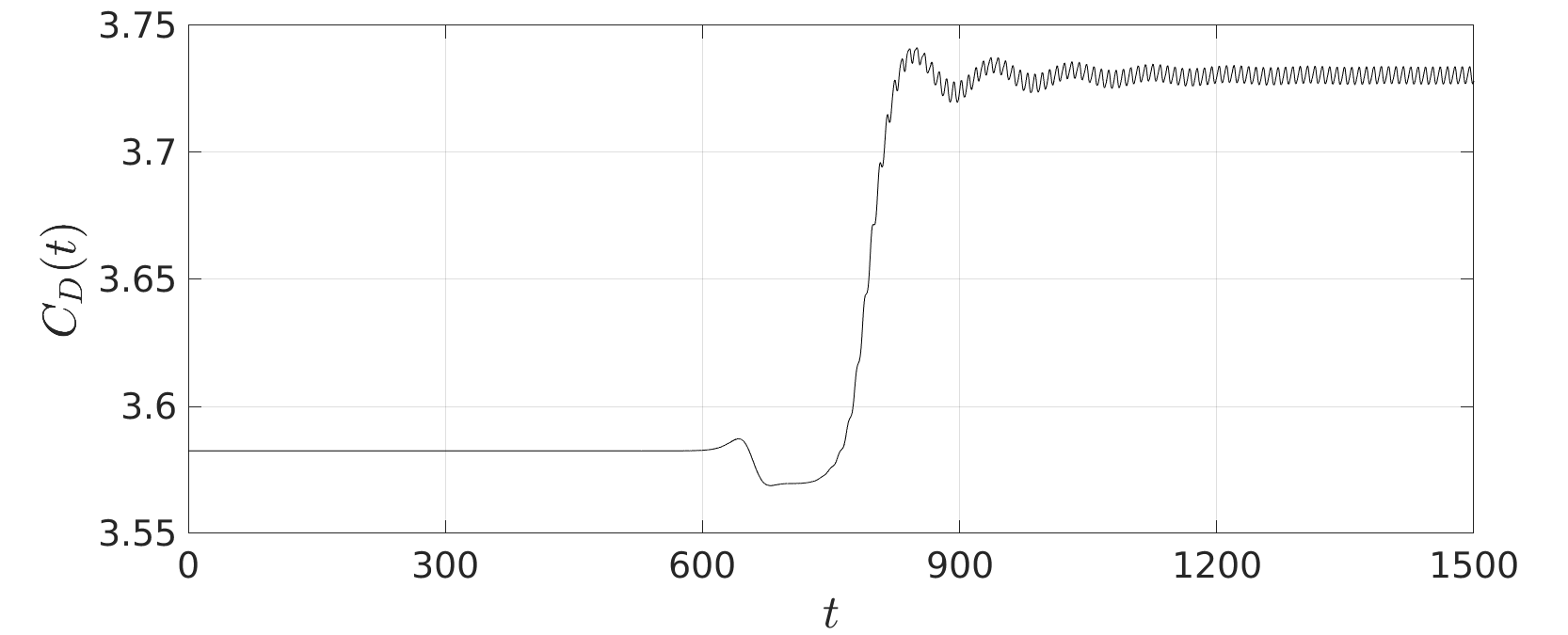}} & \raisebox{-0.92\height}{\includegraphics[width=.45\linewidth]{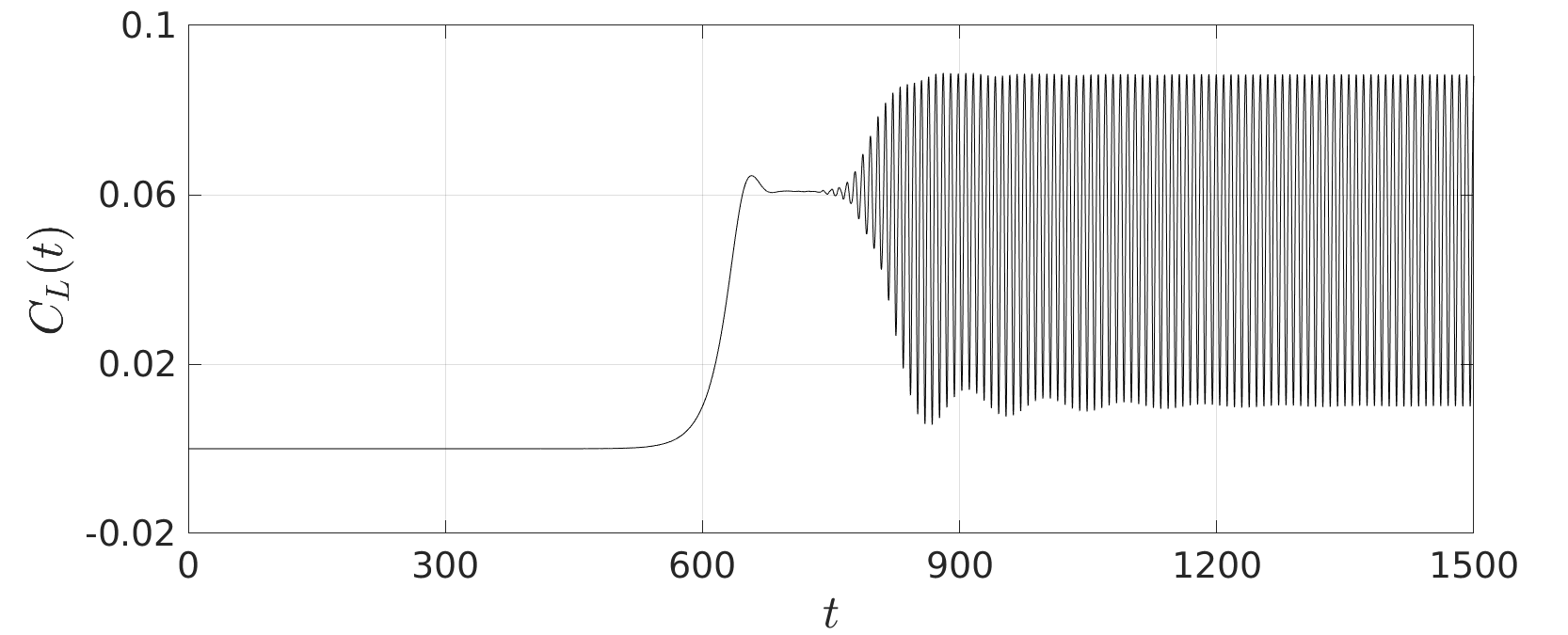}} \\ \vspace*{-3mm}
 105 & \raisebox{-0.92\height}{\includegraphics[width=.45\linewidth]{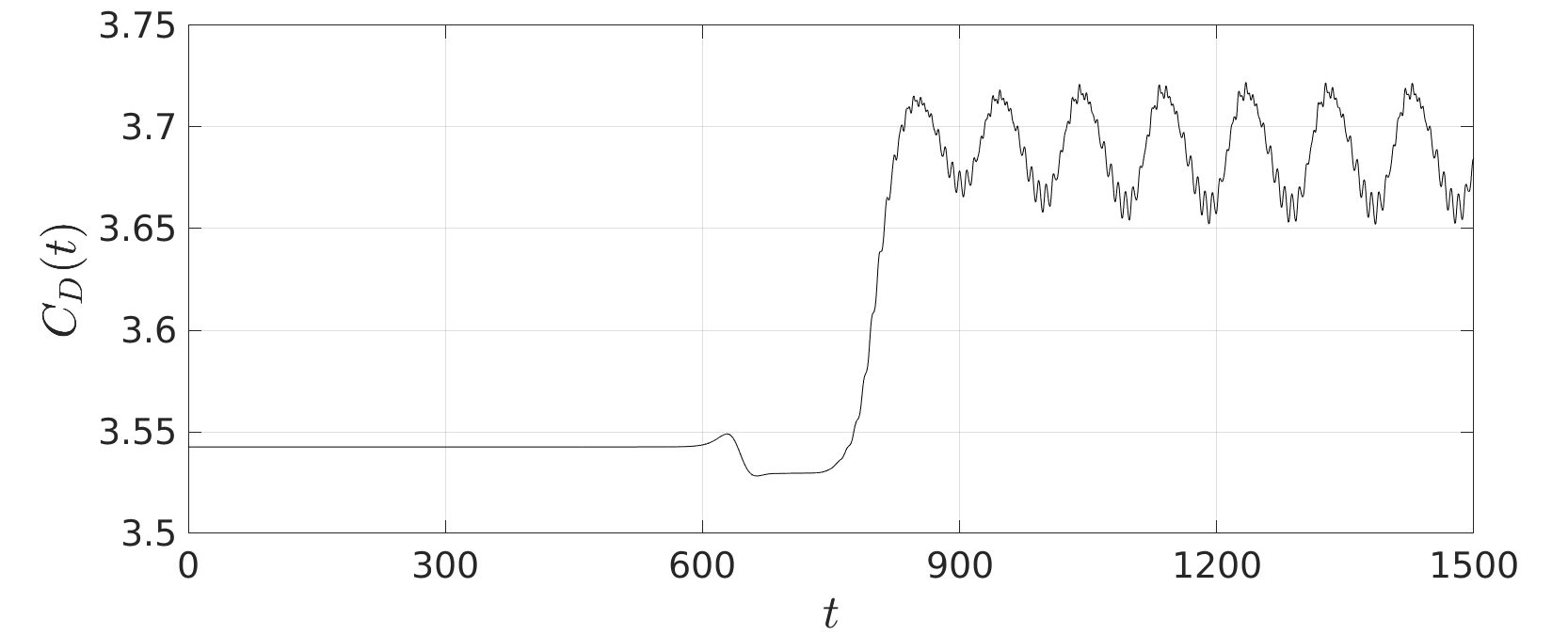}} & \raisebox{-0.92\height}{\includegraphics[width=.45\linewidth]{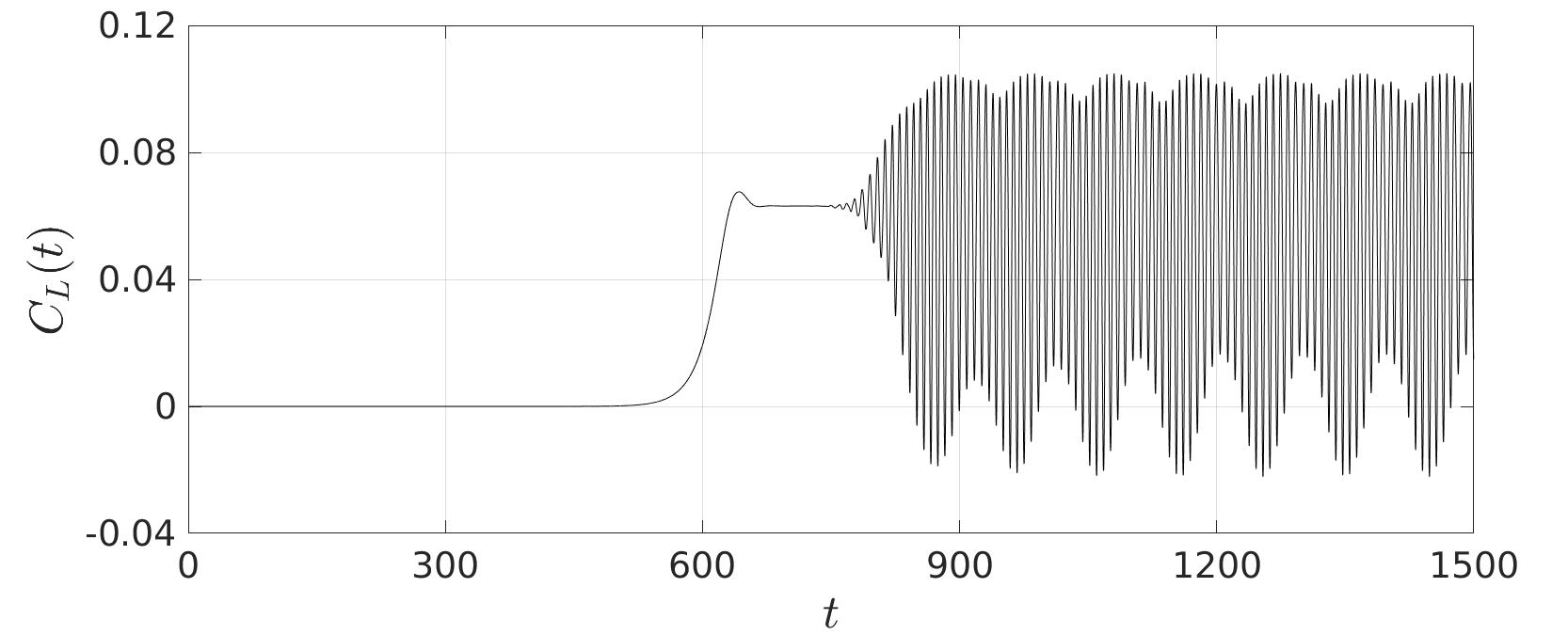}} \\ \vspace*{-3mm}
 110 & \raisebox{-0.92\height}{\includegraphics[width=.45\linewidth]{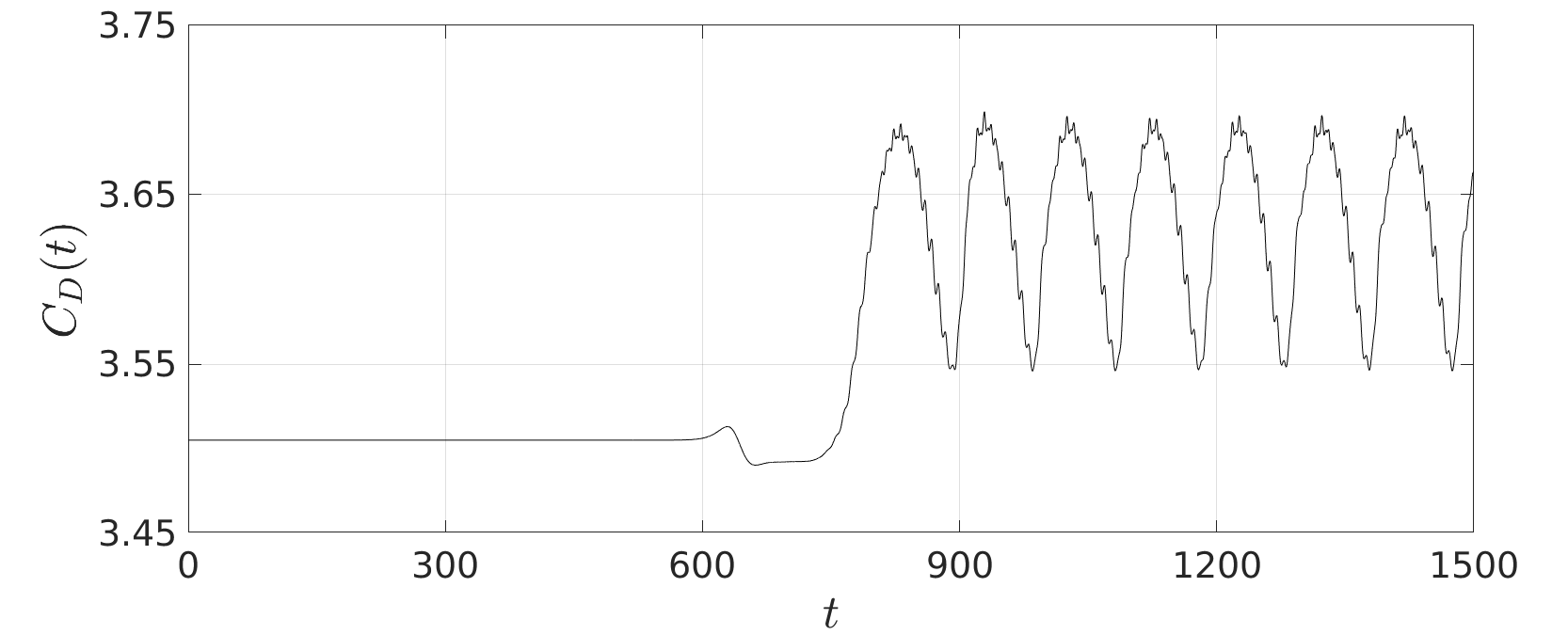}} & \raisebox{-0.92\height}{\includegraphics[width=.45\linewidth]{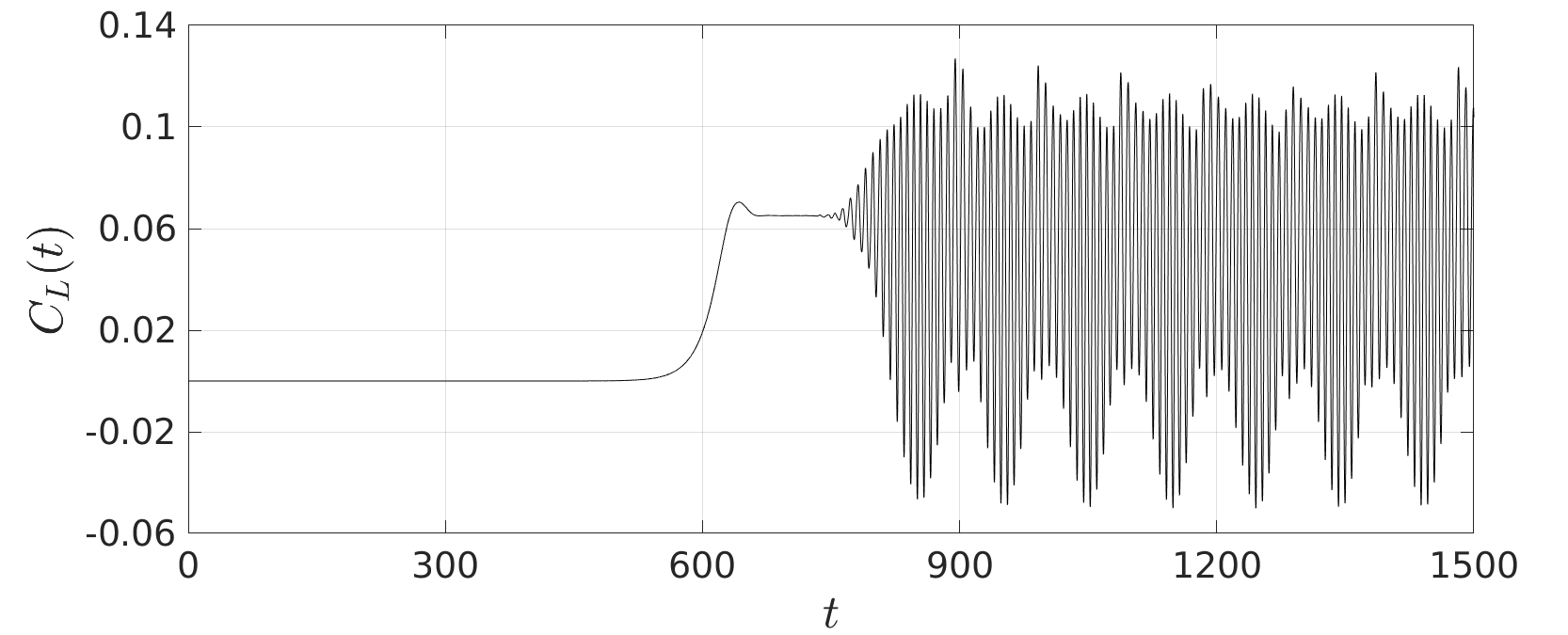}} \\ \vspace*{-3mm}
 120 & \raisebox{-0.92\height}{\includegraphics[width=.45\linewidth]{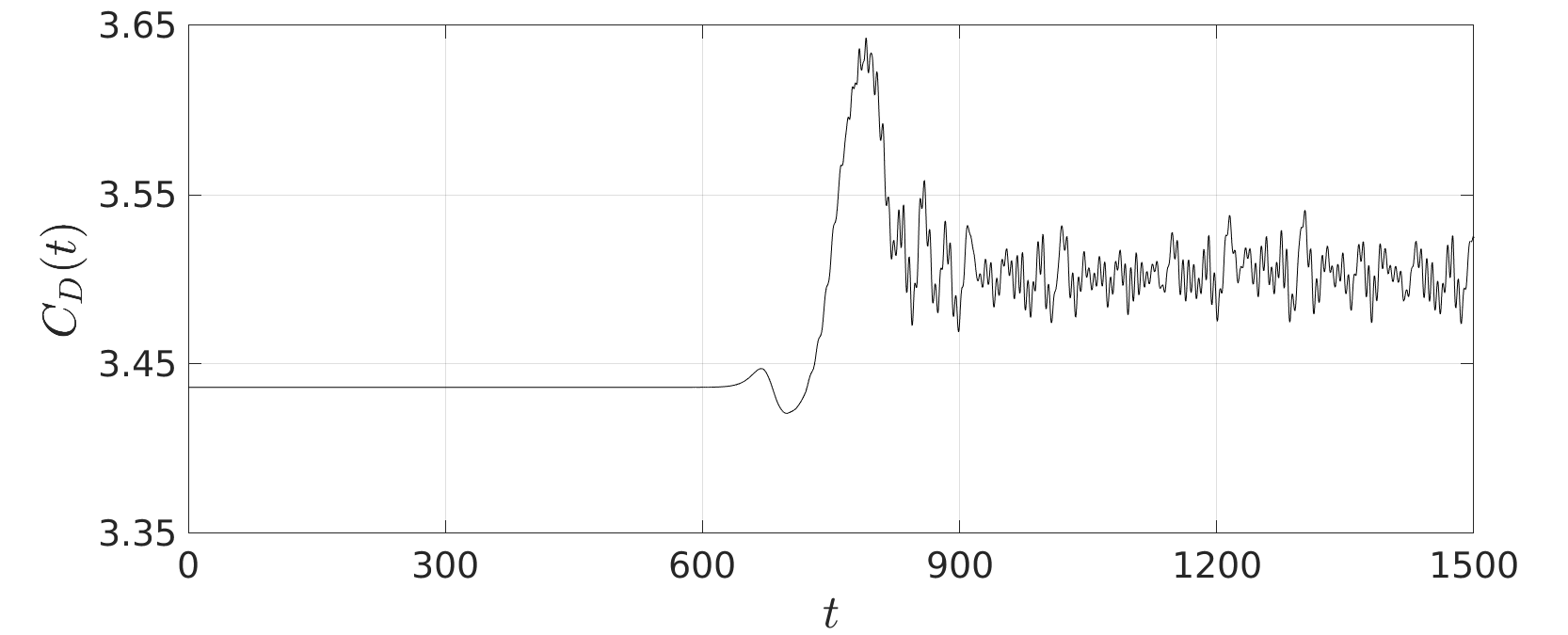}} & \raisebox{-0.92\height}{\includegraphics[width=.45\linewidth]{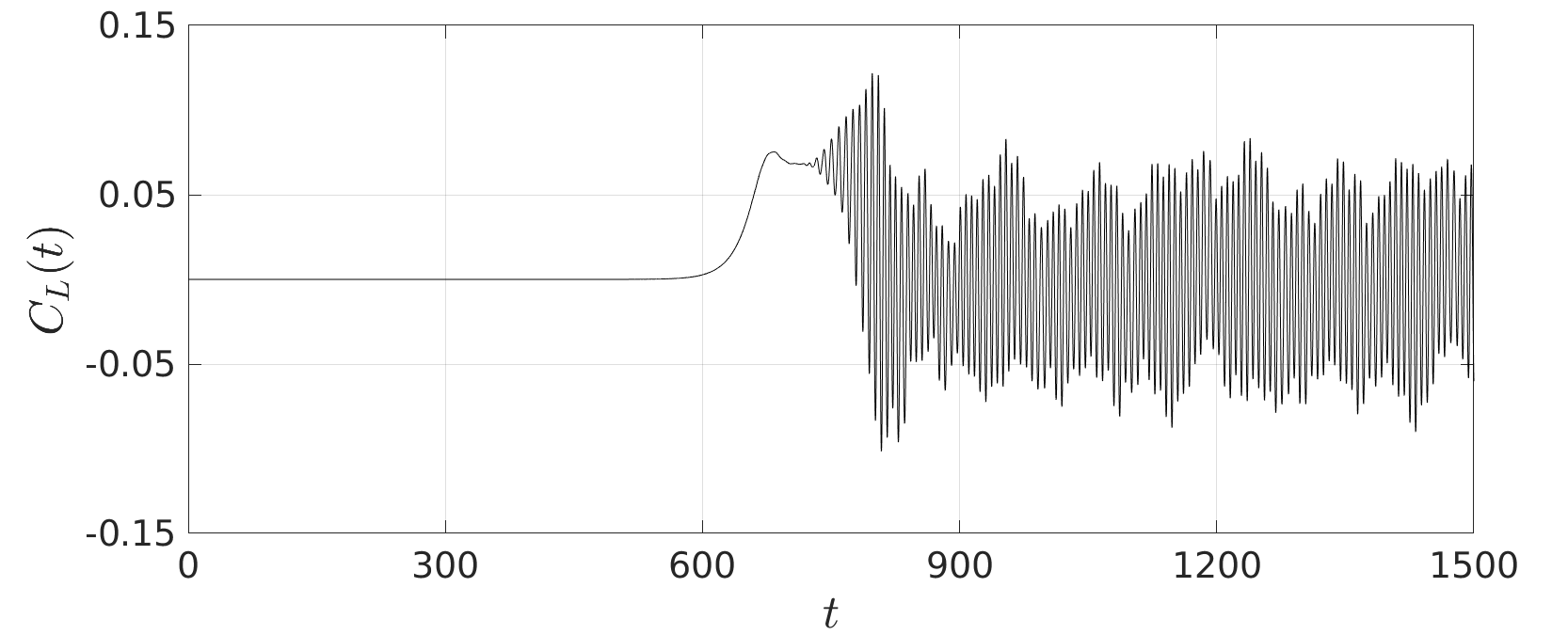}} \\
 130 & \raisebox{-0.92\height}{\includegraphics[width=.45\linewidth]{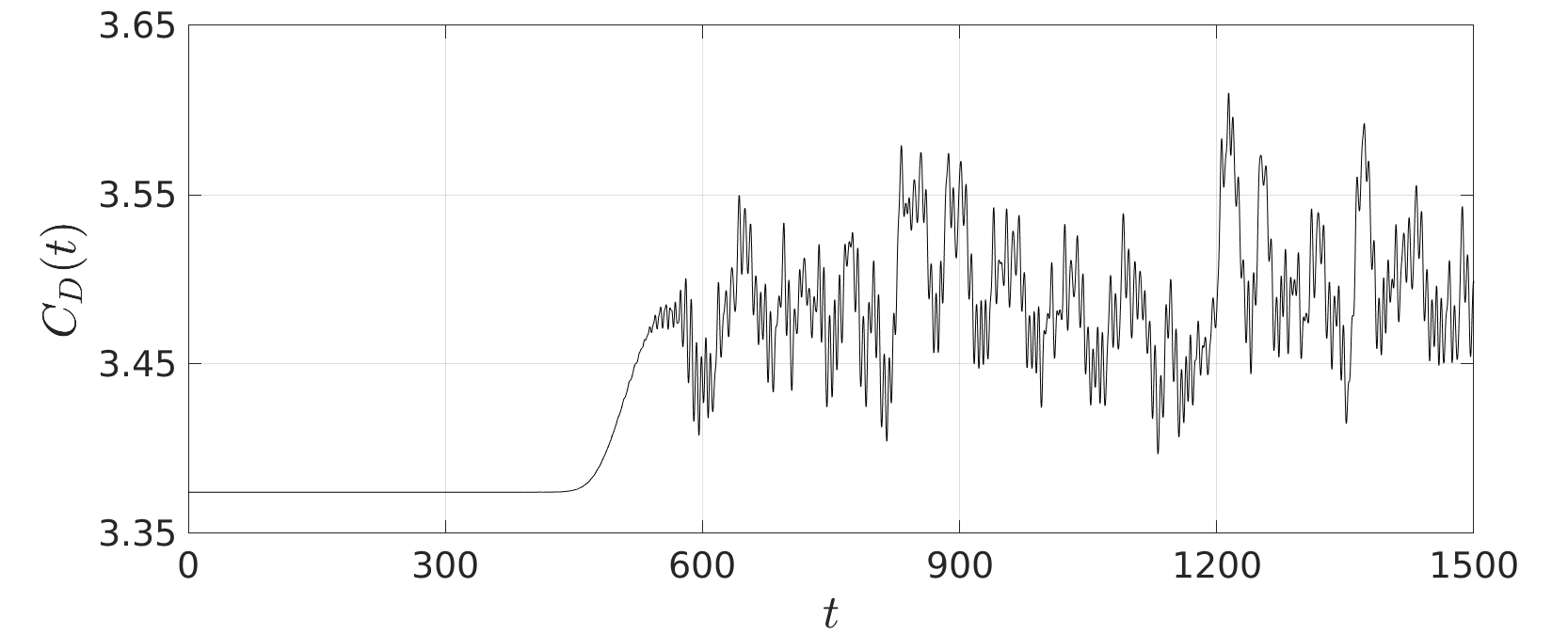}} & \raisebox{-0.92\height}{\includegraphics[width=.45\linewidth]{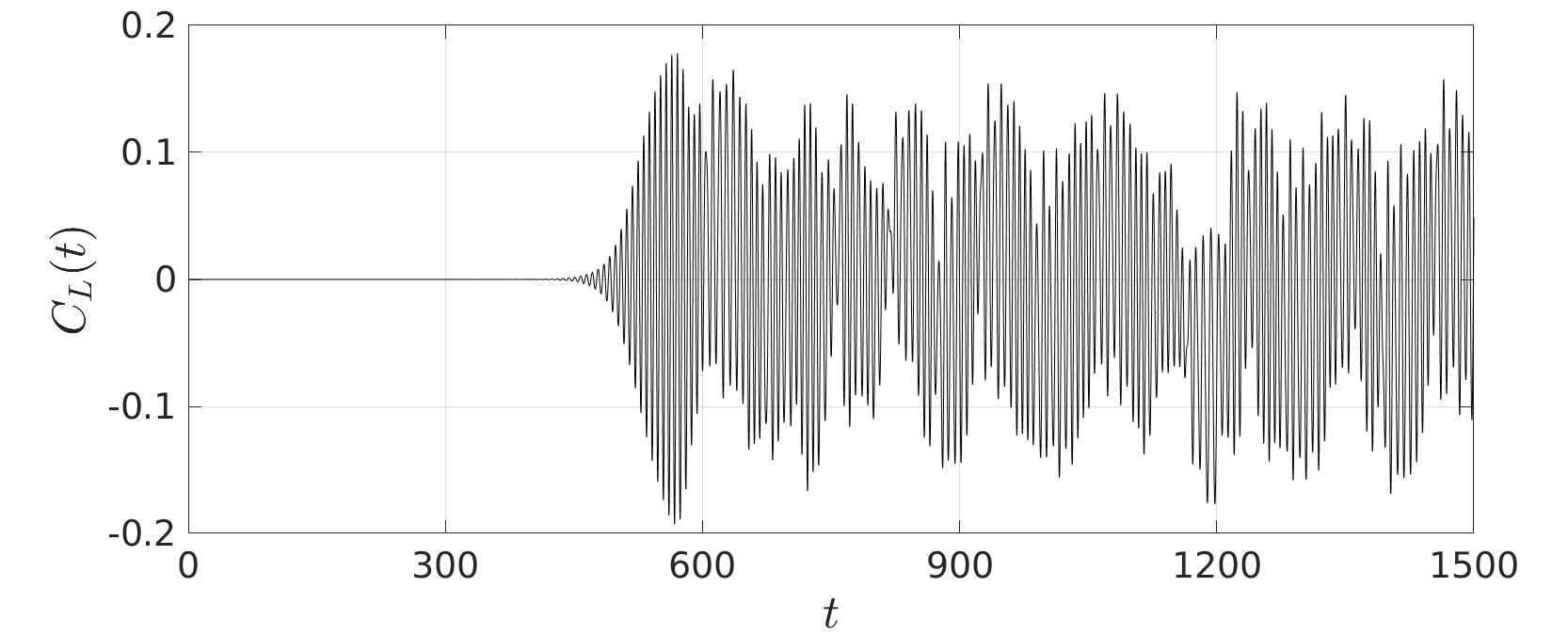}} 
\end{tabular}
}
\caption{Dynamique transitoire des coefficients de trainée $C_D$ (à gauche) et de portance $C_L$ (à droite) pour un nombre de Reynolds de valeurs croissantes. }
\label{dengfig:CLCD}
\end{figure}

Sur la figure~\ref{dengfig:CLCD} sont représentées les dynamiques transitoires des coefficients de portance $C_L$ et de trainée $C_D$ jusqu'au régime asymptotique, pour différentes valeurs du nombre de Reynolds $Re$. Dans tous les cas, la condition initiale est la solution symétrique stationnaire instable $\mathbf{u}_{0}$. Théoriquement, le système ne devrait jamais en sortir, cette solution étant un point fixe des équations. En pratique, les erreurs de discrétisation numériques rendent la solution imparfaite, et malgré un transitoire extrêmement long (plusieurs centaines de temps convectif), la trajectoire du système finit par s'en échapper, pour converger asymptotiquement vers un attracteur. La dynamique initiale démarre ainsi dans le sous-espace symétrique avec $\left\{a_1-a_5\right\}$ nuls, tandis que le régime asymptotique a quitté ce sous-espace après que les degrés de liberté $a_1$, $a_2$ et $a_4$, anti-symétriques, aient atteint leur dynamique asymptotique. On constate sur la figure~\ref{dengfig:CLCD} les résultats remarquables suivants: \\

\begin{itemize}
 \item Pour $Re=80$, lors de la dynamique transitoire, les degrés de liberté $a_{1,2,3}$ de la bifurcation de Hopf s'expriment \textit{avant} les degrés de liberté $a_{4,5}$ de la bifurcation fourche. On remarque que l'état transitoire observé sur la plage $t\in [700,800]$, symétrique en moyenne, est caractérisé par un coefficient de trainée \textit{inférieur} à celui du régime final, mais supérieur à celui de l'état initial. \\
 \item Pour les valeurs du nombre de Reynolds de 100 à 120, qui couvrent les régimes périodique ($Re=100$), quasi-périodique ($Re=105$ \& 110) et chaotique ($Re=120$), les degrés de liberté $a_{1,2,3}$ de la bifurcation de Hopf s'expriment \textit{après} les degrés de liberté $a_{4,5}$ de la bifurcation fourche. A la différence de ce qui était observé pour $Re=80$, la trainée associée à l'état stationnaire non symétrique observé autour de $t\approx 700$ pour $Re=100$ à 120, est  \textit{minimale}, tandis que la trainée associé à l'état final est maximale. \\
 \item Comme pour $Re=80$, à $Re=130$, dans le régime pleinement chaotique, les degrés de liberté $a_{1,2,3}$ s'expriment à nouveau \textit{avant} les degrés de liberté $a_{4,5}$. Ainsi, jusqu'à $t\approx 600$, le coefficient de portance reste centré ($a_4=0$), avant de devenir non nul en moyenne au-delà de ce temps ($a_4\neq 0$). \\

 \item Les oscillations de la force de trainée, pour $Re=80$ à 110, sont essentiellement dues aux modulations basse fréquence de l'écoulement. Le lâcher tourbillonnaire, en tant que tel, ne contribue que marginalement aux fluctuations asymptotiques de la trainée, contrairement à la force de portance, qui explore de grandes variations à la fréquence du lâcher tourbillonnaire. \\
\end{itemize}

Si les différents degrés de liberté agissent de manière similaire sur le coefficient de trainée pour toutes valeurs du nombre de Reynolds dans la gamme considérée, cela signifie que $a_4$ et $a_5$ ont pour effet de \textit{réduire} la trainée du sillage et que $a_3$ y contribue fortement, tandis que $a_{1,2}$ agissent essentiellement à travers l'amplitude $a_1^2+a_2^2$, et dans une moindre mesure que $a_3$, hormis en présence de modulations aux basses fréquences. Dans la section suivante, nous nous intéressons aux états d'écoulements engendrés par l'expression de ces degrés de liberté individuels.

\section{Etats transitoirement explorés}

\begin{figure}
\centerline{
\begin{tabular}{cccc}
 $Re$ & $\mathbf{u}_0$ & $\mathbf{u}(t=700)$ &  $\bar{\mathbf{u}}(t_\infty )$\\
 \hline \hline
 80 & \raisebox{-0.9\height}{\includegraphics[width=.3\linewidth]{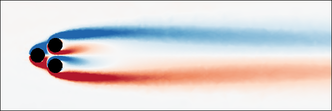}} &  \raisebox{-0.9\height}{\includegraphics[width=.3\linewidth]{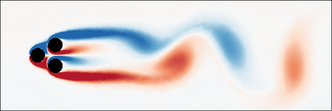}} &  \raisebox{-0.9\height}{\includegraphics[width=.3\linewidth]{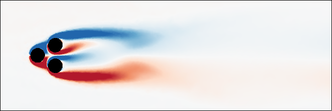}} \\
100 & \raisebox{-0.9\height}{\includegraphics[width=.3\linewidth]{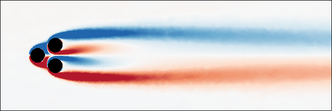}} &  \raisebox{-0.9\height}{\includegraphics[width=.3\linewidth]{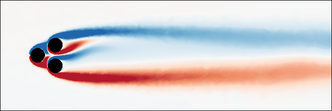}} &  \raisebox{-0.9\height}{\includegraphics[width=.3\linewidth]{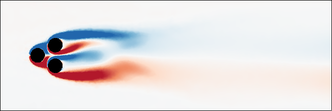}} \\
\end{tabular}
}
\caption{Etats transitoirement explorés par la dynamique du système à différentes valeurs du nombre de Reynolds: condition initiale $\mathbf{u}_0$, état au temps $t=700$, et champ moyen asymptotique $\bar{\mathbf{u}}(t_\infty )$. }
\label{dengfig:etats}
\end{figure}

\begin{figure}
\centerline{
 \includegraphics[width=.3\linewidth]{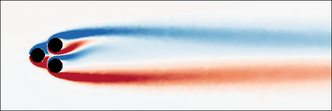} 
}
\caption{Solution stationnaire asymétrique $\mathbf{u}^+$ pour $Re=100$, à comparer au champ $\mathbf{u}(t=700)$ pour $Re=100$ de la figure~\ref{dengfig:etats}.}
\label{dengfig:u+}
\end{figure}

Sur la figure~\ref{dengfig:etats} sont représentés les états $\mathbf{u}(\mathbf{x},t=0)$ et $\mathbf{u}(\mathbf{x},t=700)$ transitoirement visités lors des dynamiques précédemment décrites, ainsi que le champ asymptotique moyen $\bar{\mathbf{u}}(\mathbf{x})$ lorsque $t\rightarrow t_\infty $. On constate que pour $100<Re<120$, le comportement transitoire de $C_L$ et $C_D$ reste constant un laps de temps non négligeable autour de $t\approx 700$, après que l'écoulement ait brisé la symétrie de la configuration géométrique, révélée ici par $C_L\neq 0$ (cf figure~\ref{dengfig:CLCD}). L'état observé sur cette plage temporelle se trouve dans un voisinage proche de la solution stationnaire asymétrique $\mathbf{u}^+$ (instable) du système, comme on peut le voir pour $Re=100$ en comparant le champ à $t=700$ sur la figure~\ref{dengfig:etats} et la solution stationnaire asymétrique $\mathbf{u}^+(\mathbf{x})$ de la figure~\ref{dengfig:u+}. Cette remarque reste valable pour $Re=105$ et $110$.
%
%
Comme on peut le voir sur la figure~\ref{dengfig:etats}, la bulle de recirculation des états $\mathbf{u}_0(\mathbf{x})$ et  $\mathbf{u}^+(\mathbf{x})$ est très étendue, ce qui explique la faible trainée associée. 
%
%
L'asymétrie des champs $\mathbf{u}(\mathbf{x},t=700)$, aux nombres de Reynolds 100 à 120, explique pourquoi le coefficient de portance cesse d'être nul en moyenne. L'action du degré de liberté $a_3$, en distordant les états de base  $\mathbf{u}_0(\mathbf{x})$ ou $\mathbf{u}^+(\mathbf{x})$ en  $\bar{\mathbf{u}}(\mathbf{x})$, réduit considérablement la zone de recirculation du sillage en aval des cylindres et contribue à l'augmentation notable de la trainée, comme cela a été évoqué dans la section précédente (voir également \cite{deng:noack2003} à propos de l'action de $a_3$ sur l'état de base). 

\section{Discussion et conclusion}

Sur la plage $[0,Re_{FC}]$ du nombre de Reynolds, seul l'état stationnaire symétrique $\mathbf{u}_0(\mathbf{x})$ est point fixe du système, stable pour $Re\leq Re_{HP}$, instable pour $Re>Re_{HP}$. Les dynamiques transitoires et asymptotiques sont alors bien décrites par l'ansatz $\mathbf{u}(\mathbf{x},t)\approx \mathbf{u}_0(\mathbf{x})+a_1(t)\mathbf{u}_1(\mathbf{x})+a_2(t)\mathbf{u}_2(\mathbf{x})+a_3(t)\mathbf{u}_3(\mathbf{x})$, où les coefficients $a_1$ à $a_3$ obéissent aux équations \eqref{dengeq:a1} à \eqref{dengeq:a3} \cite{deng:chaos2018,deng:noack2003}. Au-delà de cette valeur critique, deux nouvelles solutions stationnaires instables asymétriques, $\mathbf{u}^\pm(\mathbf{x})$, apparaissent, en plus de la solution $\mathbf{u}_0(\mathbf{x})$, comme résultat de la bifurcation fourche. Ces nouvelles solutions font intervenir les deux nouveaux degrés de liberté $a_4,a_5$, de telle sorte que $\mathbf{u}^\pm(\mathbf{x}) = \mathbf{u}_0(\mathbf{x})\pm \ a_4^{Re}\mathbf{u}_4(\mathbf{x})+a_5^{Re}\mathbf{u}_5(\mathbf{x})$, où $a_4^{Re}$ et $a_5^{Re}$ sont des valeurs des coefficients $a_4$ et $a_5$ spécifiques aux nombres de Reynolds $Re$ considérés \cite{deng:jfm2019}.

Les dynamiques transitoires aux différents nombres de Reynolds, explorées sur la route vers le chaos, révèlent des chemins empruntés dans l'espace des états du système qui évoluent avec le nombre de Reynolds. 
%
%
La force de trainée se révèle minimale lorsque les degrés de liberté $a_1,a_2,a_3$, associés à la bifurcation de Hopf, sont absents, tandis que des efforts sur les cylindres de moyenne non nulle, transverses à l'écoulement, sont présents lorsque les degrés de liberté $a_4,a_5$, associés à la bifurcation fourche, sont non nuls. 

En termes de contrôle de l'écoulement, la recherche d'une trainée minimale consisterait à supprimer l'expression des degrés de liberté $a_1,a_2,a_3$. L'écoulement résultant serait alors non symétrique aux valeurs du nombre de Reynolds considérées dans ce travail. Un écoulement symétrique de trainée minimal impliquerait la suppression supplémentaire des degrés de liberté $a_4,a_5$.\\
\newline
\textbf{Remericiements:} Ce projet est soutenu par le projet ANR-ASTRID FlowCon (ANR-17-ASTR-0022), le China Scholarship Council, le Polish National Science Center (contrat No.: DEC-2011/01/B/ST8/07264) et le Polish National Center for Research and Development (contrat No. PBS3/B9/34/2015).


\end{document}